\documentstyle[preprint,aps]{revtex}
\begin{document}
\draft

\title{Infrared response of ordered polarons in layered perovskites}
\author{P. Calvani, A. Paolone, P. Dore, S. Lupi, P. Maselli, and P. G.
Medaglia} 
\address{INFM - Dipartimento di Fisica, Universit\`a di Roma ``La Sapienza'',
Piazzale A. Moro 2, I-00185 Roma, Italy}
\author{S-W. Cheong}
\address{AT\&T Bell Laboratories, Murray Hill, New Jersey 07974, U.S.A.}    
\date{\today}
\maketitle
\begin{abstract} 
We report on the infrared absorption spectra of three oxides where charged 
superlattices  have been recently observed in diffraction experiments.  In 
La$_{1.67}$Sr$_{0.33}$NiO$_4$, polaron localization 
is found to suppress the low-energy conductivity through the opening of a
gap and to split the $E_{2u}$-$A_{2u}$ vibrational manifold of the oxygen 
octahedra. Similar effects are detected in Sr$_{1.5}$La$_{0.5}$MnO$_4$ 
and in La$_2$NiO$_{4+y}$, with peculiar differences related to the type of
charge ordering. 
\end{abstract}

The layered perovskites of general formula La$_{2-x}$Sr$_x$MO$_4$, 
with M = Ni, Mn, and Cu, at $x$=0 (or at $x$=2 for M = Mn) 
are charge-transfer, antiferromagnetic insulators and have the same crystal 
structure. While the cuprate turns into 
a high-$T_c$ superconductor at a Sr doping $x$ = 0.06, 
La$_{2-x}$Sr$_x$NiO$_{4+y}$ (LSNO) becomes metallic, not superconducting, 
at $x \sim 1$, and 
Sr$_{2-x}$La$_x$MnO$_4$ (SLMO) remains insulating at any $x$. 
Increasing effort is aimed at understanding how so different transport 
properties are related to the excitation spectrum of these materials.
The optical conductivity of the perovskites has been 
extensively discussed in the literature since the discovery of
high-$T_c$ superconductivity. Chemical doping is known to cause
the insurgence of midinfrared bands (MIR) which, according to several authors,
are responsible for the well-known "anomalous Drude" behavior
in the metallic phase.\cite{Timusk} However, the origin of such midinfrared
absorption is controversial. In cuprates like La$_{2-x}$Sr$_x$CuO$_4$
and Nd$_{2-x}$Ce$_x$CuO$_4$, the MIR band is made of two
components: i) a broad band peaked at $\sim$ 4000 cm$^{-1}$, independent of 
$T$ and generally attributed to the electronic states produced by doping in the
charge-transfer gap;\cite{Uchida91} ii) a component peaked at $\sim$ 1000 
cm$^{-1}$, strongly dependent on temperature and showing phonon-like 
structures.\cite{prb96} The latter feature has been attributed to hopping 
polarons by some authors,\cite{Falck,prb96,Yagil} to magnetic excitations by
others.\cite{Thomas92} MIR bands were also observed in 
La$_{2-x}$Sr$_x$NiO$_{4+y}$ for various $x$ and $y$, and attributed to
polarons.\cite{Bi} However, the calculated small-polaron optical conductivity 
$\sigma_{pol}$ was then found unable to describe the spectrum   
of La$_{1.84}$Sr$_{0.16}$NiO$_{4+y}$ at different temperatures.\cite{Crandles}

An attempt to define the controversy on the polaronic nature of the above 
absorption features is made in the present paper by studying three layered
perovskites where the existence of selftrapped charges has been established 
in diffraction experiments. Indeed, the photoexcitation of such charges
will produce itinerant polarons whose characteristic infrared bands 
can be first studied here in a well defined environment. We also report on 
farinfrared observations which allow to identify the lattice
deformations involved in the self-trapping of the charges. 

Recently, evidence for polaronic stripes in the basal $a-b$ plane below 
a charge ordering transition temperature $T_0\simeq$ 240 K in 
LSNO,\cite{Chen} and strong indications for bipolaronic ordered 
structures below $T_0\simeq$ 220 K in Sr$_{1.5}$La$_{0.5}$MnO$_4$,\cite{Bao} 
have been obtained by 
electron diffraction. Neutron scattering experiments on La$_2$NiO$_{4+y}$ 
(LNO)\cite{Tranquada} and Sr$_{1.5}$La$_{0.5}$MnO$_4$\cite{Sternlieb}
have shown that charge and magnetic order are strictly
related, as the polaron stripes act as charged walls which 
separate different antiferromagnetic domains. The formation of these walls 
at critical doping levels may lead to a suppression of the superconductivity, 
as observed in La$_{2-x-0.4}$Nd$_{0.4}$Sr$_x$CuO$_4$ at 
$x \sim 1/over{8}$.\cite{Tranquada2}. 

Polycrystalline  La$_2$NiO$_{4.2}$, La$_{1.67}$Sr$_{0.33}$NiO$_4$ and 
Sr$_{1.5}$La$_{0.5}$MnO$_4$, obtained as described in
Refs. \onlinecite{Chen,Bao}, have been finely milled, diluted in CsI (1:100 
in weight), and pressed in pellets under vacuum. The intensity 
$I_{s}$  transmitted by the pellet containing the oxide and
that, $I_{CsI}$, transmitted by a pure CsI pellet, were measured at the same 
$T$ by rapid scanning interferometers between 130 and 12000 cm$^{-1}$, 
after mounting both pellets on the cold finger of a
closed-cycle cryostat. A normalized optical density 
defined as $O_d(\omega)$ = ln$[I_{CsI}(\omega)/I_s(\omega)]$ was thus
obtained. In general, the optical density of a pellet will reflect 
only qualitatively the optical conductivity of the material. However, in the
present case the reflectivity of the pellet is low, the perovskite dilution 
is high, and for the dielectric function of the oxide one has
$\epsilon_1(\omega)>>\epsilon_2(\omega)$. Under these conditions, it can 
be shown\cite{Ruscher,unpub} that the above defined  
$O_d(\omega)$ is proportional to $\sigma (\omega)$, the optical conductivity of 
the pure perovskite, over the frequency range of interest here.

$O_d(\omega)$ is shown in Fig. 1 at five different temperatures for the 
polycrystalline sample La$_{2-x}$Sr$_x$NiO$_{4+y}$ with x=0.33, a  
powder where a charge/lattice modulation along the diagonals of the 
Ni-O squares on the $a-b$ planes, with a period 3${\sqrt 2} a$, was 
observed below 240 K.\cite{Chen} The spectrum at 300 K in Fig. 1 
shows a broad continuum of states with phonon peaks superimposed, which extends 
from the lowest frequencies through the charge-transfer (CT) gap, 
which is peaked at 10450 cm$^{-1}$. 
On the opposite side, $O_d$ decreases at low frequency. Then,
the broad background which at high $T$ partially shields the phonons 
cannot be attributed to conventional Drude absorption, 
consistently with the low dc conductivity of this LSNO 
sample [$\rho$ (300 K) $\sim$ 1 $\Omega$cm, $\rho$ (100 K) $\sim$ 
10$^2$ $\Omega$cm].\cite{Chen} As $T$ is lowered, an energy gap opens in the
background below $\sim$ 700 cm$^{-1}$. At 100 K the absorption does not change 
any more and a well defined midinfrared band appears. The strong dependence on 
$T$ of this band up to 3000 cm$^{-1}$ or more, is suggestive of 
polaronic absorption. Photons in the midinfrared range excite 
adiabatic polaron hopping from a perturbed site to neighboring, unperturbed 
sites.\cite{Reik} The transition rate is peaked approximately at twice the 
polaron binding energy $E_p$.\cite{Emin} The broadening at $\omega < 2E_p$ is 
due to hopping with simultaneuous annihilation of optical phonons, that at 
$\omega > 2E_p$ to hopping with creation of phonons. In some cuprates 
such vibronic lines have been resolved at low doping.\cite{prb96} 
At room temperature, the strong absorption at $\omega < 2E_p$ in 
Fig. 1 shows that the initial phonon states at high energies are populated. 
Thus, the energy needed for the charge to jump from site to site is
comparable with that of thermal excitations, and polarons are mobile. At 
lower temperatures, a depopulation of the highest phonon states opens 
the energy gap shown in Fig. 1 and prevents thermally 
activated hopping, so that the charges remain selftrapped. Then, 
the infrared spectra of Fig. 1 provide evidence for polaron localization 
between 250 and 200 K in LSNO. This does not necessarily imply that 
polarons are ordered below those temperatures. Evidence for an 
ordering transition will be discussed further on, in connection with the 
far-infrared spectrum.  
  
As already done in previous infrared investigations of the LSNO system,    
we have also compared our data with the model 
optical conductivity for a small polaron,\cite{Emin} 

\begin{equation}
\sigma_{pol} \propto (1/\omega \Delta) sinh (4E_p \omega/ \Delta^2)
exp[-(\omega^2+4E_p^2)/\Delta^2]
\end{equation}

\noindent
where $\omega$ is the photon energy and $\Delta = 2 \sqrt{2E_pE_{vib}}$. 
One may put $E_{vib}$ = (1/2) $\omega^*$ in the low-$T$ limit, 
where $\omega^*$ is a characteristic
phonon frequency, and $E_{vib} \sim kT$ in the high-$T$
limit. All quantities are expressed in cm$^{-1}$. The total optical
conductivity will be given by\cite{prb96}: 
   
\begin{equation}
\sigma (\omega) =  \sigma_{ph} (\omega) + \sigma_{pol} (\omega) +
\sigma_{MIR} (\omega) + \sigma_{CT} (\omega) 
\end{equation}

\noindent
where Lorentzian lineshapes multiplied by ($\omega/4 \pi$) are used 
for the phonon contribution $\sigma_{ph} (\omega)$, for the 
$T$-independent part of the midinfrared absorption  $\sigma_{MIR} (\omega)$,  
and for the charge-transfer band $\sigma_{CT} (\omega)$.\cite{prb96} 
For the present samples, as reported above,
$O_d(\omega) \propto \sigma (\omega)$ and one can fit Eqs. (1) and (2) 
to data in Fig. 1. The resulting curves at 300 and 20 K are plotted in 
Fig. 1 (b).  
The 300 K fit can be shown to be selfconsistent, as it yields independently 
the peak energy 2$E_p$ = 1945 cm$^{-1}$ and the bandwidth $\Delta$ = 1233 
cm$^{-1}$. Using the high-$T$ limit $\Delta = 2\sqrt {2E_p kT}$, one gets 
2$E_p$ = 1900 cm$^{-1}$ in excellent agreement with the above determination. 
The 20 K fit in turn yields 2$E_p$ = 2000 cm$^{-1}$ and, from the low-$T$ 
expression $\Delta = 2\sqrt {E_p \omega^*}$ = 1050 cm$^{-1}$, one
finds $\omega^*$ = 275 cm$^{-1}$. Here, both $\Delta$ and $\omega^*$ are
likely to be overestimated. This is due to the misfit near the 
band edge, which could possibly be eliminated by a model where at least two 
different modes are involved in polaron formation. 

The three resolved $E_{2u}$ modes of 
the $a-b$ plane, out of the four predicted by theory,  are found 
at 20 K (300 K) at 175, 235, 695 (670) cm$^{-1}$. Two $A_{2u}$ modes 
polarized along the $c$-axis, out of the three predicted, are observed 
at 300 cm$^{-1}$ as a weak shoulder, and at 501 (501) cm$^{-1}$ as a well 
resolved peak. The above assignment is based on previous 
determinations\cite{Gervais,Pintschovius} in La$_2$NiO$_4$ single crystals. 
As the temperature is lowered, major changes are observed in the phonon 
spectrum of Fig. 1. The $E_{2u}$ stretching mode at 670 cm$^{-1}$ hardens 
regularly as the lattice contracts, an effect already observed,\cite{Bi} 
while both low-energy $E_{2u}$ modes, which are shielded by
the polaronic continuum above 250 K, spring up at lower $T$'s. 

The most interesting effect of temperature involves the absorption feature 
around 350 cm$^{-1}$ (see inset). This peak includes the double
degenerate $E_{2u}$ bending mode of the in-plane Ni-O bond, and the 
$A_{2u}$ mode which displaces the 4 basal oxygens relative to the Ni 
atom and to both apical oxygens. Those two phonons, observed by neutron 
scattering in La$_2$NiO$_4$ at 347 and 346 cm$^{-1}$, 
respectively,\cite{Pintschovius} are quasi-degenerate due to the high 
symmetry of the oxygen octahedron which surrounds the Ni atom.\cite{Bates}
Any distortion of the oxygen square, as produced by a selftrapped charge, 
will remove the twofold degeneracy along the $a$ and $b$ axes of the E$_{2u}$ 
phonon, any axial distortion will remove the degeneracy between A$_{2u}$ and 
E$_{2u}$.\cite{nota1} This process is clearly monitored in the inset of Fig. 1.
Above 250 K, due to disorder and to polaron mobility, random splittings result 
in a broad absorption. At 200 K a sharp doublet appears with peak frequencies 
at 341 and 382 cm$^{-1}$, while the band narrows. This abrupt change 
yields evidence for spatial ordering of the octahedra distortions.\cite{nota2}
Indeed, no structural (i.e. tetrahedral-orthorombic) transition would
produce such a large splitting in the bending mode (see the results reported 
in Ref.\onlinecite{Gao} for La$_{2-x}$Sr$_x$CuO$_4$.) Moreover, the 
occurrence of structural transitions in {\it doped} La$_2$NiO$_4$ has been 
explicitly excluded by high-resolution neutron 
diffraction.\cite{Pintschovius} 

The Sr-free compound La$_{2}$NiO$_{4+y}$ (LNO) exhibits incommensurate 
charge and magnetic ordering in neutron scattering experiments.
At $y$=0.125 the polaronic structures are related to the ordering of the 
out-of-plane oxygens.\cite{Tranquada} Superlattice peaks which increase 
in intensity by a factor of 4 below 230 K\cite{Pintschovius} have 
been also observed in a sample with $y$=-0.13.  
These results show that incommensurate charge ordering
may occur in LNO for various oxygen nonstoichiometries. 
The absorption spectrum of La$_2$NiO$_{4.2}$ is 
reported in Fig. 2. Polaronic features similar to those of LSNO  
are observed therein, even if important differences are found. 
In Fig. 2, a gap in the polaronic continuum is partially open 
at room temperature. It deepens as $T$ lowers, before stabilizing at 100 K. 
A "freezing" in the polaronic peaks was observed at the same temperature by 
neutron scattering\cite{Pintschovius}. One may notice that in Fig. 2 
the gap opens more gradually with temperature than 
in LSNO, and that no midinfrared peak can be 
clearly identified. At 20 K (300 K) the phonon spectrum includes $E_{2u}$ 
peaks at 165, 235 (230), 665 (650) cm$^{-1}$, and $A_{2u}$ 
peaks at 300 (300), 505 (505) cm$^{-1}$. The $A_{2u}$-$E_{2u}$ manifold
at $\sim$ 350 cm$^{-1}$ (see the inset of Fig. 2) exhibits an inhomogeneuous 
broadening above 200 K, suggestive of random octahedra
deformations. At lower $T$ a doublet can be identified, much less resolved 
than the one associated with the ordering of charges in Fig. 1. 

Finally, we report in Fig. 3 the optical density of 
Sr$_{1.5}$La$_{0.5}$MnO$_4$, a sample where charge/magnetic 
superlattices similar to those of La$_{1.67}$Sr$_{0.33}$NiO$_{4}$ 
have been detected by electron diffraction\cite{Bao} and by neutron
scattering.\cite{Sternlieb} In SLMO, however, the superlattices are 
attributed to small bipolarons.\cite{Bao} 
The spectra of Fig. 3, which to our knowledge represent the first infrared
observations in the family of strontium manganate, show that here the  
charge-lattice coupling is much stronger than in LSNO. At 300 K, 
a deep absorption minimum is already present at 750 cm$^{-1}$, pointing toward 
low polaron mobilities even at high temperatures.  However, as $T$ is lowered,
a loss of spectral weight is observed below 4000 cm$^{-1}$, 
the energy range which in LSNO corresponds to 
(single) polaron absorption. At low temperature 
($T<$ 100 K) a peak is left at $\sim$ 4000 cm$^{-1}$, namely at twice the 
energy of the corresponding peak in Fig. 1. According to the Holstein model, 
if $E'_p$ is the polaron binding energy measured in SLMO, and $U<2E'_p$ is 
the repulsion energy 
between two charges on the same site, the infrared absorption 
of small bipolarons is expected to be peaked at $4E'_p - U$.\cite{Emin} 
As for LSNO $E_p \simeq$ 1000 cm$^{-1}$, $4E'_p - U \simeq 4E_p$,
so that $E'_p > E_p$ in agreement with the above statement that the
charge-phonon coupling in SLMO is larger than in LSNO. 

The assignment of the phonon peaks observed in Fig. 3 can be easily 
obtained by a comparison with the isostructural powder of Fig.
1. The TO normal modes polarized in the $a-b$ plane of 
Sr$_{1.5}$La$_{0.5}$MnO$_4$ at 20 K (300 K) are resolved 
at 190 (190), 230 (230), 645 (625) cm$^{-1}$, those along the $c$-axis 
at 285 (280) and 510 (505) cm$^{-1}$. Both lowest-energy $E_{2u}$ phonons 
are here resolved at room temperature, due to the absence of a polaronic
background at such frequencies. The Mn-O stretching mode shifts considerably
with temperature (+20 cm$^{-1}$ between 300 and 20 K) with no remarkable 
discontinuities. Here again, the ordering process of the charges detected by
diffraction probes is monitored by splittings 
in the intense E$_{2u}$-A$_{2u}$ manifold. At room 
temperature, a broad absorption peak with a few shoulders is observed,  
suggestive of disordered octahedra deformations. Between 250 
and 200 K, consistently with the bipolaron ordering transition 
at $\sim$ 220 K reported by Bao {\it et al.},\cite{Bao} three  
lines appear at 346, 388, and 433 cm$^{-1}$. The detection of a 
triplet indicates that two charges on a single site 
remove both the twofold degeneracy of the Mn-O bending mode and the accidental 
degeneracy between this mode and the A$_{2u}$ vibration. In this system, a full 
breaking of the octahedral symmetry is then produced at the sites where 
charges localize.

In conclusion, we have shown that the infrared spectra of three layered
perovskites are fully consistent with the observation in the same systems,
by diffraction probes, of charged superlattices. The reported presence of 
polarons (bipolarons) in
Ni-based (Mn-based) perovskites reflects into a partial (full) splitting of
the $E_{2u}$-$A_{2u}$ manifold related to the three-dimensional symmetry
of the oxygen octahedra. Below the temperatures where the polarons (bipolarons) 
build up superlattices, the broad distribution of 
splittings collapses into a well resolved doublet (triplet). The self-trapped
charges can be photoexcited, thus producing a midinfrared band 
that is reasonably fitted by a small-polaron optical conductivity. 
In LSNO the band is peaked at $\sim$ 2000 cm$^{-1}$, in 
SLMO at $\sim$ 4000 cm$^{-1}$. The genesis of the band is here observed. It 
forms through the opening of a gap in the polaronic background, which clears 
up the far-infrared range as polarons localize at low temperature. 
In bipolaronic Sr$_{1.5}$La$_{0.5}$MnO$_4$ the gap is already open at 300 K, 
consistently with the stronger charge-lattice coupling predicted for this
system. 

The present results may also help to interpret the  
spectra of the Cu-based perovskites, where midinfrared bands 
similar to those reported here are observed at low doping. These bands have two
components, of which that at lower frequency ($\sim$ 1000 cm$^{-1}$) has been
associated with polaron hopping by different authors. This band survives 
in the metallic phase, where it is superimposed to a Drude-like 
absorption,\cite{prb96} and can be related to the periodic charged distortions 
detected in several cuprates by extended x-ray absorption fine 
structure.\cite{Bianconiprl}

\acknowledgments
We wish to thank M. Capizzi, D. Emin, and S. Ciuchi for many 
valuable discussions. This work has been supported in part by the 
HC\&M programme of the European Union under the contract 94-0551.
 


\begin{figure}
\caption{(a) The optical density $O_d$  of polycrystalline 
La$_{1.67}$Sr$_{0.33}$NiO$_{4}$ at five temperatures
from 300 K (dotted line) to 20 K (solid line). The region of  
the  E$_{2u}$ - A$_{2u}$ manifold (see text)  
is reported in the inset by using the same symbols.  
(b)  The data at 300 and 20 K (solid lines) are fitted by use of Eqs. (1) 
and (2) (dashed lines).} 
\label{fig1}
\end{figure} 

\begin{figure}
\caption{The optical density $O_d$ of polycrystalline La$_2$NiO$_{4.2}$  
at five temperatures from 300 K (dotted line) to 20 K (solid line). 
The phonon assignment is as in Fig. 1. The region of the  E$_{2u}$ - A$_{2u}$ 
manifold (see text) is reported by using the same symbols in the inset, where 
the curves have been slightly scaled vertically for the reader's convenience} 
\label{fig2}
\end{figure} 

\begin{figure}
\caption{The optical density $O_d$ of polycrystalline 
Sr$_{1.5}$La$_{0.5}$MnO$_4$ at five temperatures from 300 K (dotted line) to 
20 K (solid line). The region of the  E$_{2u}$ - A$_{2u}$ manifold (see text)  
is reported in the inset by using the same symbols.} 
\label{fig3}
\end{figure} 


\begin{references}
\bibitem{nota2} 
   According to a model proposed to explain neutron scattering results in 
   La$_{1.9}$NiO$_{3.87}$, the octahedra would be alternately expanded and 
   contracted. Those octahedra which are expanded in the $a-b$ plane are 
   shortened along the $c$ axis, and vice versa.\cite{Pintschovius}
\end{references}
\end{document}